\documentclass{bmcart}

\usepackage[utf8]{inputenc} 
\usepackage{graphicx}

\startlocaldefs
\endlocaldefs

\begin{document}

\begin{frontmatter}

\begin{fmbox}
\dochead{Research}


\title{User's Reaction Patterns in Online Social Network Communities}


\author[
   addressref={aff1, aff2},                   
   corref={aff1},                       
   email={azza.bouleimen@supsi.ch}   
]{\inits{AB}\fnm{Azza} \snm{Bouleimen}}
\author[
   addressref={aff2},
   email={pagan@ifi.uzh.ch}
]{\inits{NP}\fnm{Nicolò} \snm{Pagan}}
\author[
   addressref={aff3},
   email={stefano.cresci@iit.cnr.it}
]{\inits{SC}\fnm{Stefano} \snm{Cresci}}
\author[
   addressref={aff2},
   email={urman@ifi.uzh.ch}
]{\inits{AU}\fnm{Aleksandra} \snm{Urman}}
\author[
   addressref={aff1},
   email={gianluca.nogara@supsi.ch}
]{\inits{GN}\fnm{Gianluca} \snm{Nogara}}
\author[
   addressref={aff1},
   email={silvia.giordano@supsi.ch}
]{\inits{SG}\fnm{Silvia} \snm{Giordano}}


\address[id=aff1]{
  \orgname{Department of Innovative Technologies, University of Applied Sciences and Arts of Southern Switzerland (DTI, SUPSI)}, 
  \city{Lugano},                              
  \cny{CH}                                    
}
\address[id=aff2]{%
  \orgname{Institute of Informatics, University of Zurich (IfI, UZH)},
  \city{Zürich},
  \cny{CH}
}
\address[id=aff3]{%
  \orgname{Institute for Informatics and
Telematics, National Research
Council (IIT, CNR)},
  \city{Pisa},
  \cny{Italy}
}


\end{fmbox}








%

\end{frontmatter}



\section*{Introduction}
Misinformation, hate speech, toxicity, trolling, and malicious bots are examples of undesired behavior on Online Social Networks (OSNs) with the potential for serious implications in real life for individuals and societies. These harmful impacts range from escalating physical violence in protests \cite{gallacher2021online}, threatening democratic elections' integrity \cite{luceri2019evolution}, 
to 
leading to genocides \cite{yue2019weaponization}.
To address these harms, OSNs platforms introduced some moderation strategies that aim at mitigating these problems. Most of these interventions follow a one-size-fits-all approach, as policies are equally applied to all users \cite{cresci2022personalized}. 
However, these interventions sometimes exacerbate the phenomena instead of limiting them \cite{horta2021platform, trujillo2022make}.
A possible explanation could consist in the fact that users present diversified reactions to moderation policies \cite{trujillo2023one}, in particular because users are typically grouped in communities within OSNs. In this context, some studies highlight the need for personalized moderation interventions \cite{cresci2022personalized}. However, to do so, we need to better understand
user's susceptibility defined as the factors that drive them toward particular reactions. In other words, we aim to get insights on what makes users more or less likely to engage in undesired behavior. 
In this work, we aim at studying the reaction of users in network communities as a first step toward understanding user's susceptibilities on the individual level. 
In turn, this represents a preliminary step towards designing personalized moderation strategies.

\section*{Results}
\textbf{Dataset.} We base our study on the VaccinItaly dataset \cite{pierri2021vaccinitaly}. It is a collection of tweets related to the Covid-19 discussion in Italy 
ranging from Dec, 20th, 2020 to Oct, 22nd 2021. The topic has been very controversial all around the world. Consequently, the choice of this dataset is suitable for studying the susceptibilities of users in contexts that prompt adversarial reactions. The dataset consists of $\sim$12 million tweets in Italian, half of which are retweets. It involves 551,816 unique users, where 86\% of them have less than 20 tweets in the dataset. We selected a subset of users that are involved enough to reflect the core discussion on the Covid-19 vaccines in Italy. To do so, we adapted and applied the definition of a \textit{core user} from \cite{trujillo2022make} to our dataset. 
This reduces the number of users to 9,278 (1.7\% of users) who are responsible for nearly half of the tweets.

The purpose of the study being to observe the behavior of users within the network structure they are in, we built the retweet network of users. It is a directed weighted network where nodes represent the users and edges represent retweets. The weights of the edges represent the number of retweets from one user to another. 

\textbf{Community detection.} We applied the Louvain community detection algorithm \cite{Blondel_2008} on the network with a resolution parameter of 0.7. We obtained two main communities that gather 87\% of the nodes in the network. We qualitatively analyzed the tweets of the nodes with the high authority scores in these two communities \cite{kleinberg1998authoritative}. In one community, the nodes with high authority scores tweet content in favor of the vaccines while, in the other community, the nodes with high authority scores are against the adoption of vaccines and the government’s measures to contain the spread of the virus. The same observations are found when analyzing the most retweeted tweets or the tweets of the most central users in these two communities. Hence, we assume that one community is dominated by a \textbf{Pro vaccination} discourse (Provax community: 3,980 nodes) and the other is dominated by an \textbf{Anti vaccination} discourse (Novax community: 3,831 nodes).

Since our work aims at measuring the differences in the reactions of users belonging to different communities, we first need to understand whether the communities were stable over time, or whether they evolved. In the latter case, differences between the two communities could be simply due to the flow of users between them. To do so, we ran different instances of the community detection algorithm on different sub-periods of time and evaluated the user's flows across communities. Our analysis showed that the composition of the two communities remains overall stable over time, hence we can use the community partitioning based on the whole dataset.


\textbf{Toxicity in communities.} In this abstract, due to space restrictions, we limit the analysis to measuring the user toxicity. Nevertheless, an analysis of negativity was also done and similar observations were obtained. To measure the toxicity of the text of the tweet, we used the Detoxify library \cite{Detoxify}. It is a state-of-the-art method for computing toxicity \cite{rossetti2023bots, maleki2021applying}. Detoxify has a multilingual model for non-English texts. For Italian it reaches an AUC of 89.18\% \cite{Detoxify}. The model returns a score ranging from 0 (low toxicity) to 1 (high toxicity). 

\begin{figure}[h!]
    \centering
    \includegraphics[width=\textwidth]{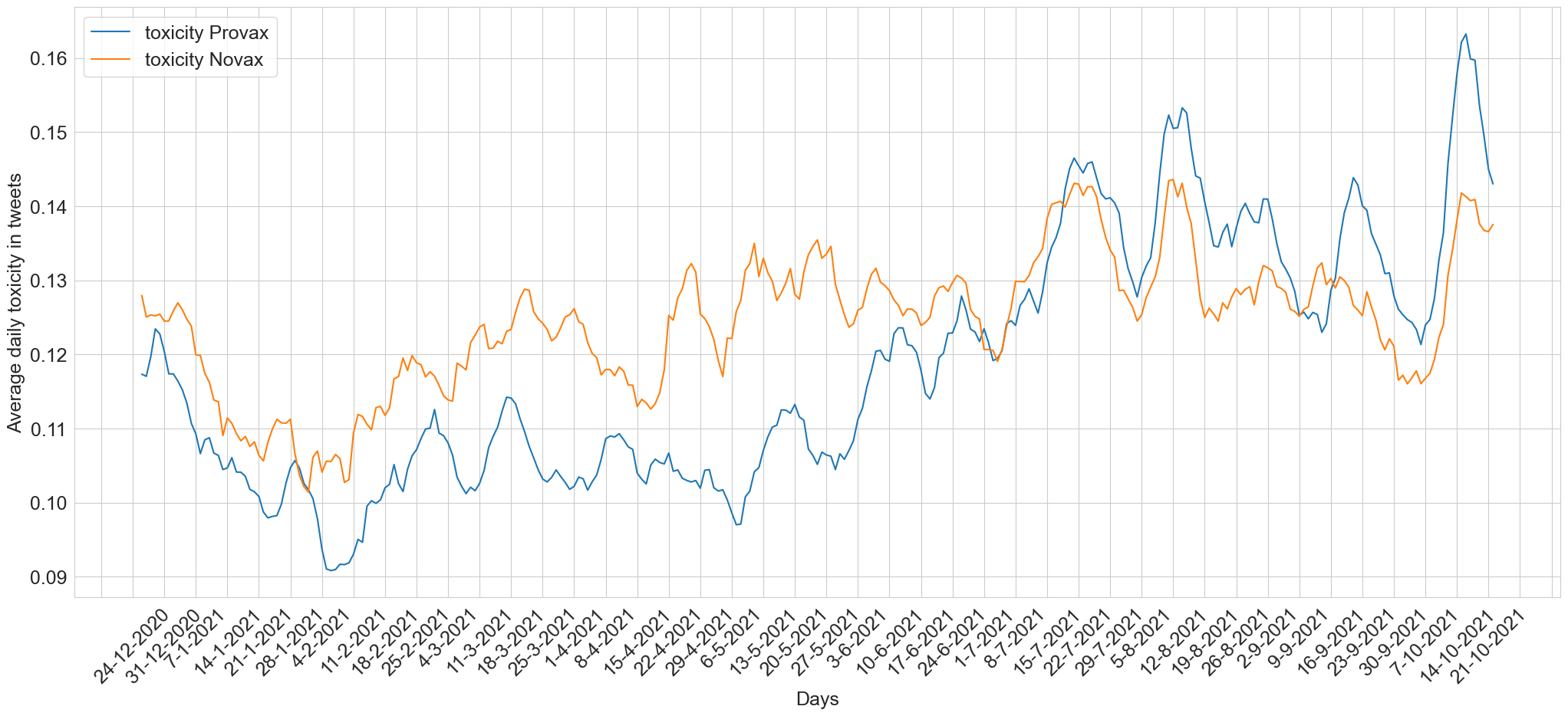}
    \caption{Daily toxicity average in written text for Provax and Novax communities.  A moving average of a 7-day window was applied to the plot.}
    \label{fig:toxicity_communities}
\end{figure}

We present in Fig. \ref{fig:toxicity_communities} the daily average toxicity of the text written by the users belonging to the Provax and Novax communities. Fig. \ref{fig:toxicity_communities} shows that, from the beginning of the data collection until mid-June, the toxicity level of the Provax community is lower than the one of the Novax community. Interestingly, this trend is inverted starting from mid-June where we notice that Provax users become on average more toxic than Novax users. Overall, the toxicity of both communities increases throughout time as shown by the Mann-Kendall test for trends. 
However, the Provax toxicity rate increases more than twice as fast as the Novax one.

\begin{figure}[h!]
    \centering
    \includegraphics[width=\textwidth]{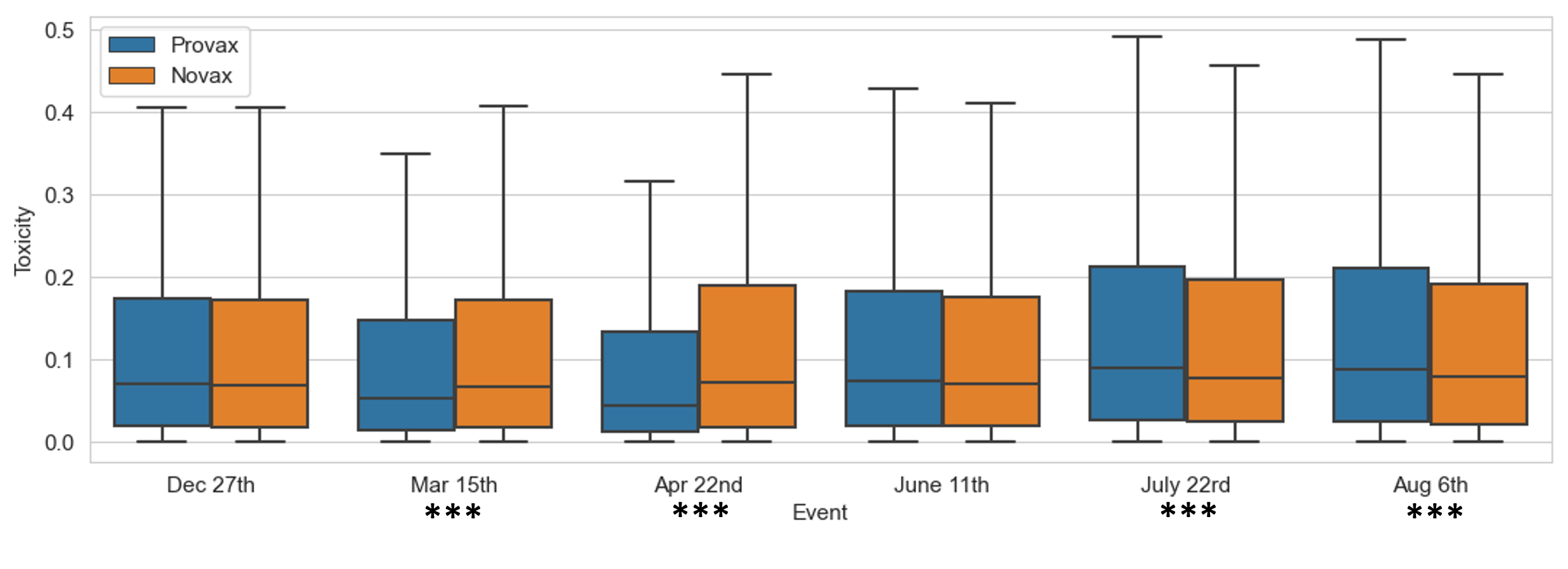}
    \caption{Box plots of the toxicity of tweets published by users of the Provax and Novax communities three days starting from the event date. The stars under the event date on the x-axis represent the significance of the difference between the two communities.}
    \label{fig:toxicity_events}
\end{figure}

\textbf{Toxicity around specific events.} To deepen the analysis, we look into the reaction of the Provax and Novax communities around specific events related to the Covid-19 pandemic in Italy. The selected events are presented in Tab. \ref{tab:events}.
\begin{table}[h!]
\caption{List of event related to the Covid-19 pandemic in Italy. Events highlighted in bold correspond to pics in the number of tweets posted by the user on that day.}
\label{tab:events}
\begin{tabular}{ll}
Date           & Event                                                                                                                           \\ \hline
\textbf{Dec. 27\textsuperscript{th}, 2020}  & Start of the vaccination campaign in Italy.                       \\
\textbf{Mar. 15\textsuperscript{th}, 2021}  & The Italian government announces lockdown during eater vocation.     \\
Apr. 22\textsuperscript{nd}, 2021  & Introducing the use of the Green Pass.                                                                                          \\
\textbf{Jun. 11\textsuperscript{th}, 2021} & Death of a young woman after receiving Astrazeneca shot.         \\
\textbf{Jul. 23\textsuperscript{rd}, 2021} & Announcing mandatory Green Pass starting from the 6\textsuperscript{th} of August 2021. \\
Aug. 6\textsuperscript{th}, 2021   & Mandatory Green Pass required to access several public spaces.                                                                 
\end{tabular}
\end{table}
In Fig. \ref{fig:toxicity_events}, we present box plots of the toxicity of the tweets posted by the users of the two communities for the three days following a specific event. Both Provax and Novax have a similar reaction, in terms of toxicity, to the start of the vaccination campaign on Dec 27\textsuperscript{th}. In fact, there is no significant difference between the toxicity of the two communities. For Mar. 15\textsuperscript{th} and Apr. 22\textsuperscript{nd}, the toxicity of the Novax community is significantly higher. This relates to the trends of toxicity observed in Fig. \ref{fig:toxicity_communities}. For Jun. 11\textsuperscript{th}, the difference between the two communities is not significant anymore. This happens at the time where, in Fig. \ref{fig:toxicity_communities}, we recognize an increase in the toxicity level of the Provax community reaching the level of the Novax one. For Jul. 22\textsuperscript{nd} and Aug. 6\textsuperscript{th}, it is the Provax that is this time significantly more toxic than the Novax community. Fig. \ref{fig:toxicity_events} illustrates that users belonging to different communities develop different reaction patterns depending on which events they are confronted with. It also supports the inversion in the temporal trend of the toxicity within both communities observed in Fig. \ref{fig:toxicity_communities}. This result shows that some events might trigger among groups of users a reaction that can shape the behavior of a whole community, while the impact can be non-existent for other groups of users. This supports the need for intervention strategies targeted at the group and at the individual level.


\section*{Conclusion}
Through the case study of the VaccinItaly dataset, we studied the reaction of users to specific events on OSNs considering their position in the network. We found that these events impact differently the OSN users. They can significantly alter the behavior of a community as a whole and invert the dynamics of behavior within the whole network. Our work highlights the presence of an understudied phenomenon which is the user's susceptibility to undesired behavior. It stresses out the importance of understanding the reasons behind the changes in users' reactions and fine-tuning the research to the individual's level. Possible paths forward include investigating social contagion effects, the interplay between the reactions in the two communities, and the existence of a relation between the structure of the network, the position of a user within the graph, and their reaction to a particular event. 
We hope, through our contribution, to pave the way towards building better OSNs' intervention strategies centered on the user.






\bibliographystyle{bmc-mathphys} 
\bibliography{bmc_article}      
\end{document}